\documentclass[journal,10pt]{IEEEtran}
\usepackage{graphicx, epstopdf}
\usepackage{amssymb}
\usepackage{amsmath}
\usepackage{amsthm}
\usepackage{mathrsfs}
\usepackage{cite}
\usepackage{color}
\usepackage[acronym]{glossaries}
\usepackage[T1]{fontenc}
\usepackage[latin9]{inputenc}
\usepackage{array}
\usepackage{textcomp}
\usepackage{multirow}
\usepackage{soul}

\title{
Lightwave Power Transfer for Federated Learning-based Wireless Networks
}
\author{
        {
        Ha-Vu Tran, Georges Kaddoum, Hany Elgala, Chadi Abou-Rjeily and Hemani Kaushal
        }
\thanks{
Ha-Vu Tran and Georges Kaddoum are with University of Qu\'{e}bec, \'{E}TS engineering school, LACIME Laboratory, Montreal, Canada  (e-mails: ha-vu.tran.1@ens.etsmtl.ca, georges.kaddoum@etsmtl.ca).

Ha-Vu Tran is also with Duy Tan University, Da Nang, Viet Nam.

Hany Elgala is with the Electrical and Computer Engineering Department, University at Albany-State University of New York, Albany, (e-mail: helgala@albany.edu).

Chadi Abou-Rjeily is with the Department of Electrical and Computer Engineering of the Lebanese American University (e-mail: chadi.abourjeily@lau.edu.lb).

Hemani Kaushal is with University of North Florida, Jacksonville, FL, USA, (e-mail: hemani.kaushal@unf.edu).
}
 }
\begin{document}

    \maketitle
\begin{abstract}
Federated Learning (FL) has been recently presented as a new technique for training shared machine learning models in a distributed manner while respecting data privacy. However, implementing FL in wireless networks may significantly reduce the lifetime of energy-constrained mobile devices due to their involvement in the construction of the shared learning models.  To handle this issue, we propose a novel approach at the physical layer based on the application of lightwave power transfer in the FL-based wireless network and a resource allocation scheme to manage the network's power efficiency.  Hence, we formulate the corresponding optimization problem and then propose a method to obtain the optimal solution. Numerical results reveal that, the proposed scheme can provide sufficient energy to a mobile device for performing FL tasks without using any power from its own battery. 
Hence, the proposed approach can support the FL-based wireless network to overcome the issue of limited energy in mobile devices.

\end{abstract}

\section{Introduction}
Recently, the concept of Federated Learning (FL) has been introduced by Google \cite{google,Jakub2016}. The main idea behind FL is to build a shared learning model based on data sets that reside across multiple terminal devices while protecting data privacy. In FL networks, each device computes the gradient updates based on its local training data. The updates are then sent to a central server in order to be aggregated in the current global shared model. Afterward, the central server feedbacks the new global model to all devices. By doing so, raw local training data are not leaked.

In the past few years, FL has attracted increasing attention from the research community \cite{Jakub2016,Yang2019,nguyentran2019}.  Nevertheless, the following critical issue was overlooked in the literature: in FL-based wireless networks, mobile devices are energy-constrained, hence, the energy consumed for executing FL might significantly decrease the devices' lifetime \cite{nguyentran2019}. To overcome this issue, prolonging the devices' lifetime, performed at the physical layer (PHY), can be a prominent solution. In this context, wireless power transfer (WPT) over radio frequency (RF) waves is interesting \cite{Zhang2013}. However, deploying RF WPT results in a performance trade-off between RF energy transfer and RF information transmission due to the RF spectrum scarcity \cite{Zhang2013}. Further, because of the propagation loss and the restriction of human exposure to RF signals \cite{IEEEstandard}, RF WPT may not provide wireless devices with sufficient energy for executing FL tasks, including computation and uplink transmission. Thus, this motivates researchers to seek a new WPT technique, which relies on a license-free spectrum and can provide a better performance than the RF WPT. In this regard, wireless lightwave energy recharging, operating in the visible light (VL) and infrared light (IRL) parts of the electromagnetic spectrum, has recently gained great interest from both academia and industry since it does not interfere with existing RF communication systems. Particularly, its high potential in enabling continuous wireless recharging has been confirmed in \cite{Fakidis,GaofengPan,Diamantoulakis2018,Tran2019}.

In this work, we propose the use of lightwave power transfer to enable new possibilities for the sustainability of future FL-based wireless networks. Accordingly, we consider a FL-based network scenario where each terminal device harvests energy from VL and IRL transmitted by an optical transmitter and then uses this energy for: (i) computing the gradient updates based on its local training data and (ii) conveying them to an access point (AP) via RF uplink communication. On this basis, we aim to derive a resource allocation scheme to handle the power efficiency in the network. In this regard, optimizing the transmit light power from the optical transmitter, the time-slots of computation and uplink transmission at each device and the receive beamformers at the RF AP constitute interesting open research problems. In addition, the total energy consumption for executing FL is restricted by the harvested energy while the uplink rate and the transmit power budget are constrained by preset thresholds. The resulting optimization problem is difficult to solve because multiple variables are coupled in constraints. Therefore, we propose a method to tackle the problem in an efficient way. The contributions of our work can be summarized as proposing, for the first time, the application of power transfer through light in FL-based wireless networks and deriving the optimal solution for the resulting problem.

\section{System Model}
\begin{figure}[t]
\centering
{\includegraphics[width=0.33\textwidth]{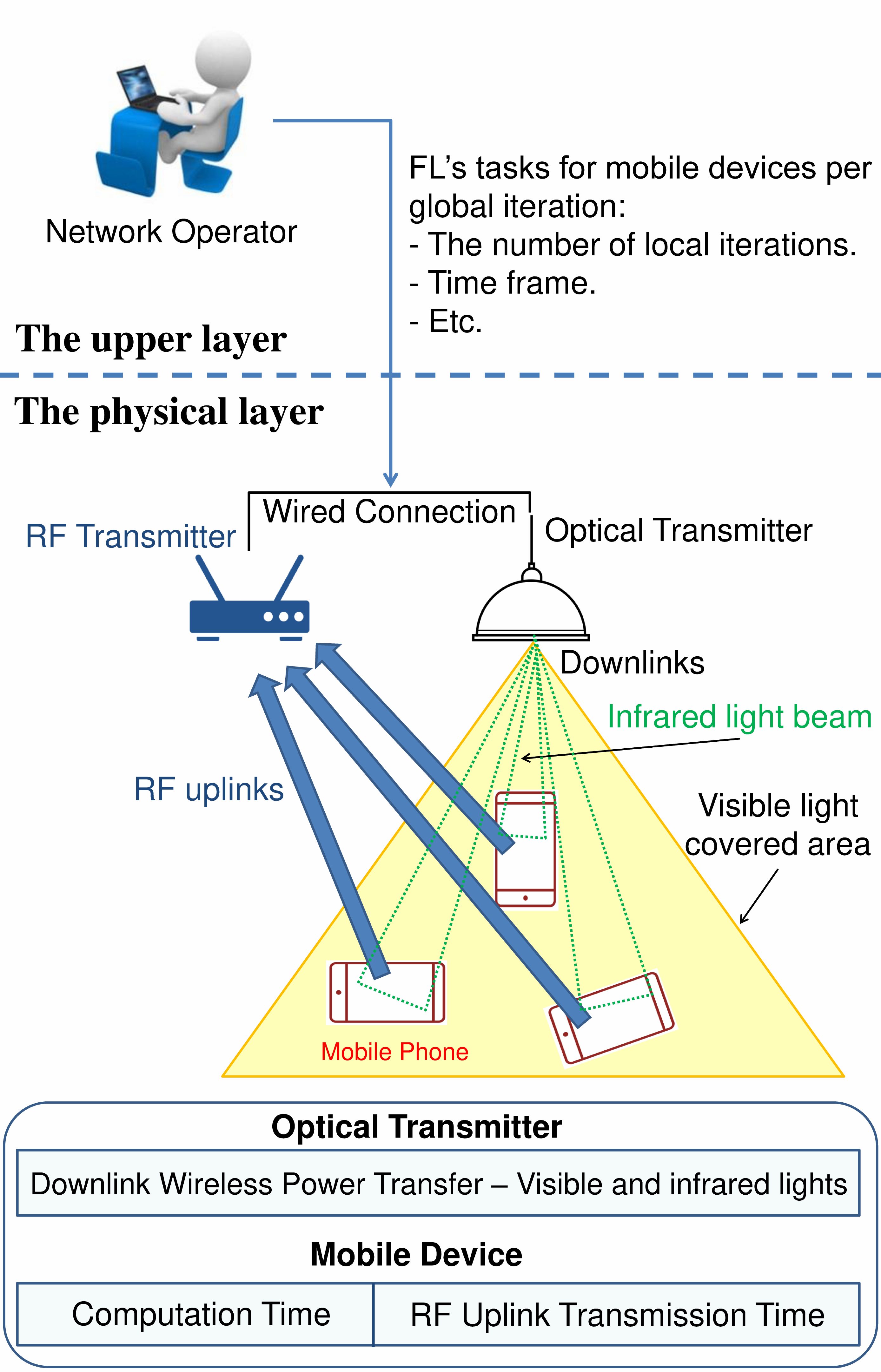}}
   \caption{
       Scenario of lightwave power transfer for FL-based networks.
    }
    \label{fig:system}
\end{figure}

Recently, the LIGHTS transmitter developed by the Wi-Charge company is able to wirelessly recharge a mobile phone.
This inspires us to consider the network model depicted in Fig. \ref{fig:system} where one optical transmitter aims to recharge $J$ terminal devices by using VL and IRL in downlink transmission while the devices communicate with an $M$-antenna RF AP through RF uplink transmissions. Each device is equipped with a single antenna, and a transparent solar panel integrated into the device's screen. We assume that a line-of-sight (LOS) transmission exists between the optical transmitter and individual devices.

Particularly, in this work, we assume that managing the network to achieve a given accuracy level within a given convergence time of the FL model is done by a network operator in upper layers. Hence, at the PHY, it is supposed that each terminal device has been assigned with FL tasks, such as the preset numbers of global and local update iterations and a preset time frame for communication and computation. On this basis, in contrast to previous works \cite{Jakub2016,Yang2019,nguyentran2019}, our letter focuses on addressing the problem of how to sufficiently supply wireless energy to terminal devices for their FL tasks at each global iteration with the preset number of local iterations and the time frame.

At the PHY, the proposed FL-based scenario is described as follows:
\begin{itemize}
\item The terminal devices harvest energy from downlink VL and IRL. This energy will be used to accomplish the FL tasks consisting of computation and RF uplink transmission.
\item Each device computes the gradient update to build a shared learning model based on its local training data. Next, the devices send their gradient updates to the RF AP over the RF uplink channels and need to finish their uplink transmissions at the same time. The gradient updates are synchronously aggregated once they are all are collected at the RF AP. 
\end{itemize}

\subsection{Optical Downlink Wireless Power Transfer}
\subsubsection{Channel Model}
We consider optical channels with only the LOS component since the contribution of a non-line-of-sight (NLOS) component to the mission of power transfer could be neglected  \cite{GaofengPan,Diamantoulakis2018}. Thus, the optical channel between the optical transmitter and the receiver photodetector of a device $j$ $(1 \le j \le J)$ is denoted by $h_{i,j}$, i.e., $i = \{ 0;1 \}$. Further, $h_{0,j}$ and $h_{1,j}$ represent the VL and IRL channels, respectively, are given by \cite{Diamantoulakis2018}:
\begin{align}
h_{i,j} = \frac{A_{j} (m_{i} + 1)}{2\pi d_{j}^2} \text{cos}^{m_{i}}(\phi_{i,j}) T_s(\psi_{i,j} ) g(\psi_{i,j} ) \text{cos}(\psi_{i,j} ), 
\end{align}
where $A_{j}$ is the photodetector's active area, $m_{i}$ is the Lambert's mode number, $d_{j}$ is the distance between the optical transmitter and device $j$, $\phi_{i,j}$ is the irradiation angle, $\psi_{i,j}$ is the angle of incidence, $T_s(\psi_{i,j} )$ is the optical band-pass filter gain, and $g(\psi_{i,j} )$ is the optical concentrator gain. Moreover, $m_{i}$ and $g(\psi_{i,j} )$ are computed according to the LED semi-angle at half-power, denoted by $\phi_{i,1/2}$, and the field of view (FOV), denoted by $\psi_{i,j,c} \le \pi/2$ \cite{GaofengPan}.

\subsubsection{Lightwave Energy Harvesting}
The optical harvested energy at device $j$ during a time frame ${\tau}$ is
\begin{align}\label{eq:totalVLCEH}
 \mathtt {EH}_{j} = \tau\sum _{i = 0,1} \mathtt {EH}_{i,j},
\end{align}
where $\mathtt {EH}_{i,j}$ can be computed as \cite{Diamantoulakis2018}, \cite{CLi2011}
\begin{align}\label{eq:VLCEH}
 \mathtt {EH}_{i,j}^{} = f_\mathrm{opt}I_{i,j,G} V_{i,j,c},
\end{align}
in which $f_\mathrm{opt}$ is the fill factor and ${I_{i,j,G}}$ is the generated DC current which can be calculated as
\begin{align}\label{eq:IG}
I_{i,j,G} = \nu P_{i,j} h_{i,j},
\end{align}
where $\nu$ is the photodetector responsivity, $P_{i,j}$ is the transmit power of the DC component, and $V_{i,j,c}$ is the open circuit voltage computed as
\begin{align}\label{eq:Voc}
V_{i,j,c} = V_t \text{ln} \left(1 + \dfrac{I_{i,j,G}}{I_d}\right),
\end{align}
where $V_t$ is the thermal voltage and $I_d$ is the dark saturation current.

Note that $\mathtt {EH}_{i,j}$ refers to the maximum obtainable energy that the solar panel can generate. Generally, a solar panel can continuously work at the maximum power point under an appropriate management strategy. Thus, this model can be used to gauge the EH performance of solar panels \cite{CLi2011}.

\subsection{FL Computation Model}
Each device keeps a local data set, denoted by $\mathcal D_j$. In the supervised learning setting, the data set $\mathcal D_j$ may include $D_j$ input-output pairs, so that the task of each device is to find model parameters which map an input to an output.
We denote the number of CPU cycles needed for each device to process one input-output pair by $c$. {In particular, the value of $c$ can be computed offline \cite{Miettinen2010}}. Thus, for a given user $j$, the CPU energy consumption to process all its data for one local iteration can be expressed as follows \cite{nguyentran2019}
\begin{align}\label{eq:Ecomp}
E^\mathrm{ comp}_j = \frac{\alpha}{2} c D_j (f^\mathrm{ CPU}_j)^2 = \frac{\alpha}{2} (c D_j)^3 \frac{1}{(T^\mathrm{ comp}_j)^2}.
\end{align}
where $\frac{\alpha}{2}$ is the effective capacitance coefficient of the device's computing chipset and $f^\mathrm{ CPU}_j$ is the CPU-cycle frequency. 
Furthermore, the corresponding computation time per local iteration of each device is defined as 
$T^\mathrm{ comp}_j = \frac{cD_j}{f^\mathrm{ CPU}_j}$.

\subsection{Radio Frequency Uplink Transmission}
We denote the uplink transmission channel between device $j$ and the RF AP by ${\mathbf g}_j \in \mathbb C ^{M}$.
Accordingly, for a given transmission time $T^\mathrm{trans}_j$, the achievable uplink data rate of device $j$ can be computed as
\begin{footnotesize}
\begin{align}\label{eq:uplinkrate}
{R_{U,j} = T^\mathrm{ trans}_j B \text {log}_2\left(1 + \frac{ \left| {  {\mathbf g}_j^H {\mathbf w}_j} \right|^2 P_{U,j}}{ {\mathbf w}_j^H \left( \sum\limits_{j' \ne j } {\mathbf g}_{j'} {\mathbf g}_{j'}^H +  \sigma^2_{0}  {\mathbf I} \right) {\mathbf w}_j }\right)},
\end{align}
\end{footnotesize}where $B$ is the bandwidth, ${\mathbf w}_j \in \mathbb C ^{M}$ denotes the receive beamforming vector at the RF AP, $P_{U,j}$ is the transmit power of the device, $\sigma^2_{0}$ is the variance of the additive white Gaussian noise (AWGN), ${\mathbf I}$ is the identity matrix and (.)$^H$ stands for the Hermitian operation.
Furthermore, the transmission energy consumption at user $j$ is
\begin{align}\label{eq:Etrans}
E^\mathrm{ trans}_{j} = T^\mathrm{ trans}_j P_{U,j}.
\end{align}

\section{Problem Formulation and Proposed Solution}

\subsection{Problem Formulation}
For the power efficiency purpose, we aim to minimize the IRL transmit power, subject to the constraints of the RF uplink rate, the total energy consumption used for computation and transmission, and the power budget. Particularly, to maintain consistent illumination, the transmit power and the light beam of VL can not be flexibly changed, and thus should not be considered as variables in the problem formulation. Hence, the corresponding optimization problem can be formulated as follows:
\begin{subequations}\label{eq:MOProblem1}
\begin{align}
	\text{OP$_1$:} \quad   \underset{{\mathbf w}_j, P_{1,j}, P_{U,j}, \atop f^\mathrm{ comp}_j, T^\mathrm{ trans}_j} \min \quad & \sum\limits_{\forall j} P_{1,j}  \label{eq:MOProblema} 
\\
\text{s.t.:}  	\quad & R_{U,j} \ge \theta_j, \quad (\forall j) \label{eq:MOProblemb}
\\			
			\quad & E^\mathrm{ comp}_j + E^\mathrm{ trans}_j \le  \mathtt {EH}_{j}, \quad (\forall j) \label{eq:MOProblemc}
\\
			\quad & 0 \le P_{1,j} \le \mathtt P_j, \quad (\forall j)	\label{eq:MOProblemg}
\\			
			\quad & K_jT^\mathrm{ comp}_j + T^\mathrm{ trans}_j = \tau, \quad (\forall j) \label{eq:MOProblemd}
\\			
			\quad & f^\mathrm{ min}_j \le f^\mathrm{ CPU}_j \le f^\mathrm{ max}_j, \quad (\forall j) \label{eq:MOProbleme}
\\			
			\quad & \left| {\mathbf w}_j \right|^2 = 1, \quad (\forall j) \label{eq:MOProblemf}					
\end{align}
\end{subequations}
where constraint \eqref{eq:MOProblemb} is set to ensure that the uplink rate of device $j$ is greater than or equal to the threshold $\theta_j$. Constraint \eqref{eq:MOProblemc} implies that the energy consumed for the computation and the uplink transmission is lower than the harvested energy from the downlink. The IRL transmit power is constrained in \eqref{eq:MOProblemg} by the power budget $\mathtt P_j$. In constraint \eqref{eq:MOProblemd}, $K_j$ is the required number of local iterations. This constraint implies that the total time for computation and transmission is equal to the time frame ${\tau}$. Note that $K_j$ and $\tau$ are thresholds set by an upper-layer. Constraint \eqref{eq:MOProbleme} imposes the CPU-frequency range of the devices.  Finally, constraint \eqref{eq:MOProblemf} implies that the receive beamforming vectors have unit power. 

It can be observed that OP$_1$ has an intractable form since variables ${\mathbf w}_j, P_{1,j}, P_{U,j}, f^\mathrm{ CPU}_j$, and $T^\mathrm{ trans}_j$ are coupled in constraints \eqref{eq:MOProblemb}, \eqref{eq:MOProblemc}, and \eqref{eq:MOProbleme}. Hence, solving OP$_1$ is challenging.

\subsection{Proposed Optimal Solution}
\subsubsection{Optimal Receive Beamformers ${\mathbf w}^{\star}_j$}
We start by observing constraints \eqref{eq:MOProblemb} and \eqref{eq:MOProblemc}. First, optimizing ${\mathbf w}_j$ does not impact optimizing ${\mathbf w}_{j'}$ $(\forall {j' \ne j })$. Second, solving problem OP$_1$ implies minimizing the sum of $E^\mathrm{ comp}_j$ and $E^\mathrm{ trans}_j$. Then, the optimal value of ${\mathbf w}_j$ is the one maximizing $R_{U,j}$ in order to {reduce $T^\mathrm{ trans}_j$ and $P_{U,j}$, following from equations \eqref{eq:uplinkrate} and \eqref{eq:Etrans}.}
In light of this discussion and \eqref{eq:MOProblemf}, the optimal value of ${\mathbf w}_j$ can be computed as
\begin{align}
{\mathbf w}^{\star}_j =  \arg \underset{\left| {\mathbf w}_j \right|^2 = 1} \max  R_{U,j} \quad (\forall j).
\end{align}
Hence, based on Rayleight-Ritz quotient \cite{Parlett}, ${\mathbf w}^{\star}_j$ is the eigenvector corresponding to the largest eigenvalue of the matrix $ {\mathbf g}_j {\mathbf g}_j^H \left( \sum_{j' \ne j } {\mathbf g}_{j'} {\mathbf g}_{j'}^H +  \sigma^2_{0}  {\mathbf I} \right)^{-1}$. 

\subsubsection{Eleminating Variables $f^\mathrm{ comp}_j$ and $T^\mathrm{ comp}_j$}
To make problem OP$_1$ more tractable, we aim to suppress variables $f^\mathrm{ comp}_j$ and $T^\mathrm{ comp}_j$ in constraints \eqref{eq:MOProblemc} and \eqref{eq:MOProbleme}.

According to \eqref{eq:Ecomp}, and \eqref{eq:MOProblemd}, $E^{\rm comp}_j$ can be derived as
\begin{align}\label{eq:Ecomp1}
E^\mathrm{ comp}_j =  \frac{\frac{\alpha}{2} (c D_j)^3 K_j^2}{\left(\tau - T^\mathrm{ trans}_j\right)^2}.
\end{align}
Next, let $\Gamma_j = \frac{ \left| {  {\mathbf g}_j^H {\mathbf w}_j^{\star}} \right|^2 }{ {\mathbf w}_j^{\star H} \left( \sum\limits_{j' \ne j } {\mathbf g}_{j'} {\mathbf g}_{j'}^H +  \sigma^2_{0}  {\mathbf I} \right) {\mathbf w}_j^{\star} }$. Hence, \eqref{eq:uplinkrate} and \eqref{eq:MOProblemb} can be rewritten as
\begin{align}\label{eq:Pu}
P_{U,j} = \frac{2^{\frac{\theta_j}{T^\mathrm{ trans}_j B}}-1}{\Gamma_j}.
\end{align}
Thus, \eqref{eq:Etrans} can be rewritten as
\begin{align}\label{eq:Etrans1}
E^\mathrm{ trans}_j = T^\mathrm{ trans}_j \frac{2^{\frac{\theta_j}{T^\mathrm{ trans}_j B}}-1}{\Gamma_j}.
\end{align}
By substituting \eqref{eq:Ecomp1} and \eqref{eq:Etrans1} into \eqref{eq:MOProblemc}, constraint \eqref{eq:MOProblemc} can be further expressed as
\begin{align}\label{eq:MOProblemc2}
\frac{\frac{\alpha}{2} (c D_j)^3 K_j^2}{\left(\tau - T^\mathrm{ trans}_j\right)^2} + T^\mathrm{ trans}_j \frac{2^{\frac{\theta_j}{T^\mathrm{ trans}_j B}}-1}{\Gamma_j} \le \tau \mathtt {EH}_{j}.
\end{align}

Furthermore, in light of \eqref{eq:Ecomp} and \eqref{eq:MOProblemd}, constraint \eqref{eq:MOProbleme} can be reformulated as
\begin{align}\label{eq:MOProbleme1}
0 < \tau-\frac{cD_jK_j}{f^\mathrm{ min}_j} \le T^\mathrm{ trans}_j \le \tau-\frac{cD_jK_j}{f^\mathrm{ max}_j} < \tau. \quad (\forall j)
\end{align}

\subsubsection{Decomposing problem OP$_1$ into subproblems without the loss of optimality}
Based on the characteristic of OP$_1$, one can observe that minimizing the sum of optical powers is equivalent to minimizing the individual ones, i.e., \{ $P_{1,j}$\}. This follows from the fact that the variables associated with each user are not coupled.

Moreover, since $\mathtt {EH}_{j}$ is an increasing function of $P_{1,j}$, this implies that $P_{1,j}$ reaches its minimum once the part on the left side of \eqref{eq:MOProblemc2}, denoted by $\Psi(T^{\rm trans}_j)$, i.e., $\Psi(T^{\rm trans}_j) = \frac{\frac{\alpha}{2} (c D_j)^3K_j^2}{\left(\tau - T^\mathrm{ trans}_j\right)^2} + T^\mathrm{ trans}_j \frac{2^{\frac{\theta_j}{T^\mathrm{ trans}_j B}}-1}{\Gamma_j}$, reaches its minimum over $T^{\rm trans}_j$. Thus, this suggests that one needs to find the minimum of $\Psi(T^{\rm trans}_j)$ and then seeks the minimum of $P_{1,j}$.

In light of the above analysis, without loss of optimality, we decompose OP$_1$ into the two following subproblems
\begin{subequations}
\begin{align}
	\text{SubOP$_{j1}$:} \quad  \underset{ T^\mathrm{ trans}_j} \min \quad & \Psi(T^\mathrm{ trans}_j)  \label{eq:SubOP21a}
\\
\text{s.t.:} 	\quad & \text{eq. } \eqref{eq:MOProbleme1}, \nonumber					
\end{align}
\end{subequations}
\begin{subequations}
\begin{align}
	\text{SubOP$_{j2}$:} \quad  \underset{ P_{1,j}} \min \quad & P_{1,j} 
\\
\text{s.t.:} 	\quad & \mathtt {EH}_{j} = \epsilon_j^{\star},  \label{eq:SubOP22b}	 \hspace{10pt}\\
			\quad & \eqref{eq:MOProblemg}, \nonumber				
\end{align}
\end{subequations}
where $\epsilon_j^{\star} = \Psi(T^\mathrm{\star trans}_j)$ and $T^\mathrm{\star trans}_j$ is obtained by solving SubOP$_{j1}$.

\subsubsection{Solving SubOP$_{j1}$ for Optimal Transmission Time $T^{\star \rm trans}_j$}
SubOP$_{j1}$ is convex and its convexity can be verified by evaluating the second derivative of $\Psi(T^\mathrm{ trans}_j)$ as follows
\begin{align}
\frac{\text{d}^2 \Psi(T^\mathrm{ trans}_j)}{\text{d}(T^\mathrm{ trans}_j)^2 } = \frac{3{\alpha} (c D_j)^3 K_j^2}{\left(1 - T^\mathrm{ trans}_j\right)^4} + \frac{\theta_j^2 2^{\frac{\theta_j}{T^\mathrm{ trans}_j B}}}{B^2(T^\mathrm{ trans}_j)^3\Gamma_j},
\end{align}
which is larger than 0 under constraint \eqref{eq:MOProbleme1}. 
Further, one can see that $T^\mathrm{trans}_j$ is the only variable in SubOP$_{j1}$ and its value is bounded by constraint \eqref{eq:MOProbleme1}.
Then, SubOP$_{j1}$ can be solved using the Golden-section search method \cite{William2007} where $T^{\rm trans}_j$ is updated until convergence with the following rule: 
\begin{center}
\begin{tabular}{|l|}
\hline
{If} $\Psi(a_{n+1}) \le  \Psi(b_{n+1})$ { then} $T^\mathrm{ trans}_j \in [ a_n, b_{n+1}]$. \tabularnewline
\quad { Else} $T^\mathrm{ trans}_j \in [ a_{n+1}, b_{n}]$. \tabularnewline
\hline
\end{tabular}
\end{center}
Herein, $a_{n+1} = a_n + \rho (b_n - a_n)$, $b_{n+1} = a_n + (1-\rho) (b_n - a_n)$, $\rho = \frac{3-\sqrt{5}}{2}$, $a_0 = \tau-\frac{cD_jK_j}{f^\mathrm{ min}_j} $, and $b_0 = \tau-\frac{cD_jK_j}{f^\mathrm{ max}_j}$ \cite{William2007}. The optimal solution is found based on consecutively narrowing the interval inside which the solution exists by using the Golden ratio $\rho$. The solution is simple to achieve and the method is guaranteed to converge.

\subsubsection{Solving SubOP$_{j2}$ for Optimal Power $P_{1,j}^{\star}$}
Constraint \eqref{eq:SubOP22b} can be rewritten as
\begin{align}
\mathtt {EH}_{1,j} = \frac{ \epsilon_j^{\star} }{\tau} - \mathtt {EH}_{0,j}. \quad (\forall j)
\end{align}
Following from \eqref{eq:VLCEH}, \eqref{eq:IG}, and \eqref{eq:Voc}, constraint \eqref{eq:SubOP22b} can be reformulated as
\begin{align}\label{eq:VLCsubProblemb1}
 \text{ln} \left(1 + \dfrac{I_{{i,j},G} (P_{1,j})}{I_d}\right) \ge \frac{ \frac{ \epsilon_j^{\star} }{\tau} - \mathtt {EH}_{0,j}}{f_\mathrm{opt} V_{t} I_{{i, j},G} (P_{1,j}) },
\end{align}
where $I_{i,j,G} (P_{1,j})$ denotes that $I_{i,j,G}$ is a function of $P_{1,j}$. 

SubOP$_{j2}$ has only one variable to be minimized, i.e., $P_{1,j}$. The value of $P_{1,j}$ is bounded by constraint \eqref{eq:MOProblemg}. Compared with SubOP$_{j1}$, SubOP$_{j2}$ has one additional constraint (i.e., \eqref{eq:VLCsubProblemb1}). 
Thus, based on the given characteristics, SubOP$_{j2}$ can be tackled by a bisection-based algorithm \cite{William2007}. By setting $P_{\rm min} =  0$ and $P_{\rm max} = \mathtt P_j$, $P_{1,j}$ is updated until convergence through the rule below:
\begin{center}
\begin{tabular}{|l|}
\hline
{If} \eqref{eq:VLCsubProblemb1} is satisfied with $P_{1,j} = \dfrac{P_\mathrm{min} + P_\mathrm{max}}{2}$, \tabularnewline
{then} $P_\mathrm{ min} = P_{1,j}$.
 {Else} $P_\mathrm{ max} =P_{1,j}$. \tabularnewline
\hline
\end{tabular}
\end{center}
Using the bisection search, the value range of the optimal solution is narrowed through repeatedly bisecting the interval according to the above rule. So this method is straightforward and obtaining the final solution is not challenging.

\section{Numerical Results}
In this simulation, we consider the environment shown in Fig. \ref{fig:system}. We assume that the three devices, namely 1, 2, and 3, are located $3.3$ m, $3$ m, and $2.7$ m away from the RF AP, respectively; while being $2.3$ m, $2.2$ m, $2.1$ m away from the optical transmitter, respectively.
Regarding the uplink RF channels, we set $M=4$. 
Further, the uplink RF channels are assumed to follow a Rician distribution with a Rician factor of {8} dB and a pathloss exponent factor equal to 2.6. 
For the optical downlink channels, important parameters are listed in Table 1.
For the uplink rate, $B = 1$ MHz, and $\sigma_0^2 = 10^{-10}$ W. 
For the light EH model, $I_d = 10^{-9}$ mA,  $f_\mathrm{opt} = 0.75$, and $\nu = 0.4$ A/W (i.e., silicon solar cell).
For the computation model, $\alpha = 2*10^{-28}$, $c = 20$, $\{D_j\} = 10$ Mb, $f^\mathrm{max}_j = 1.5$ GHz, and $f^\mathrm{min}_j = 0.3$ GHz \cite{nguyentran2019}. The simulation is carried out over 10000 channel realizations.

\begin{figure}[t]
\centering
{\includegraphics[width=0.32\textwidth]{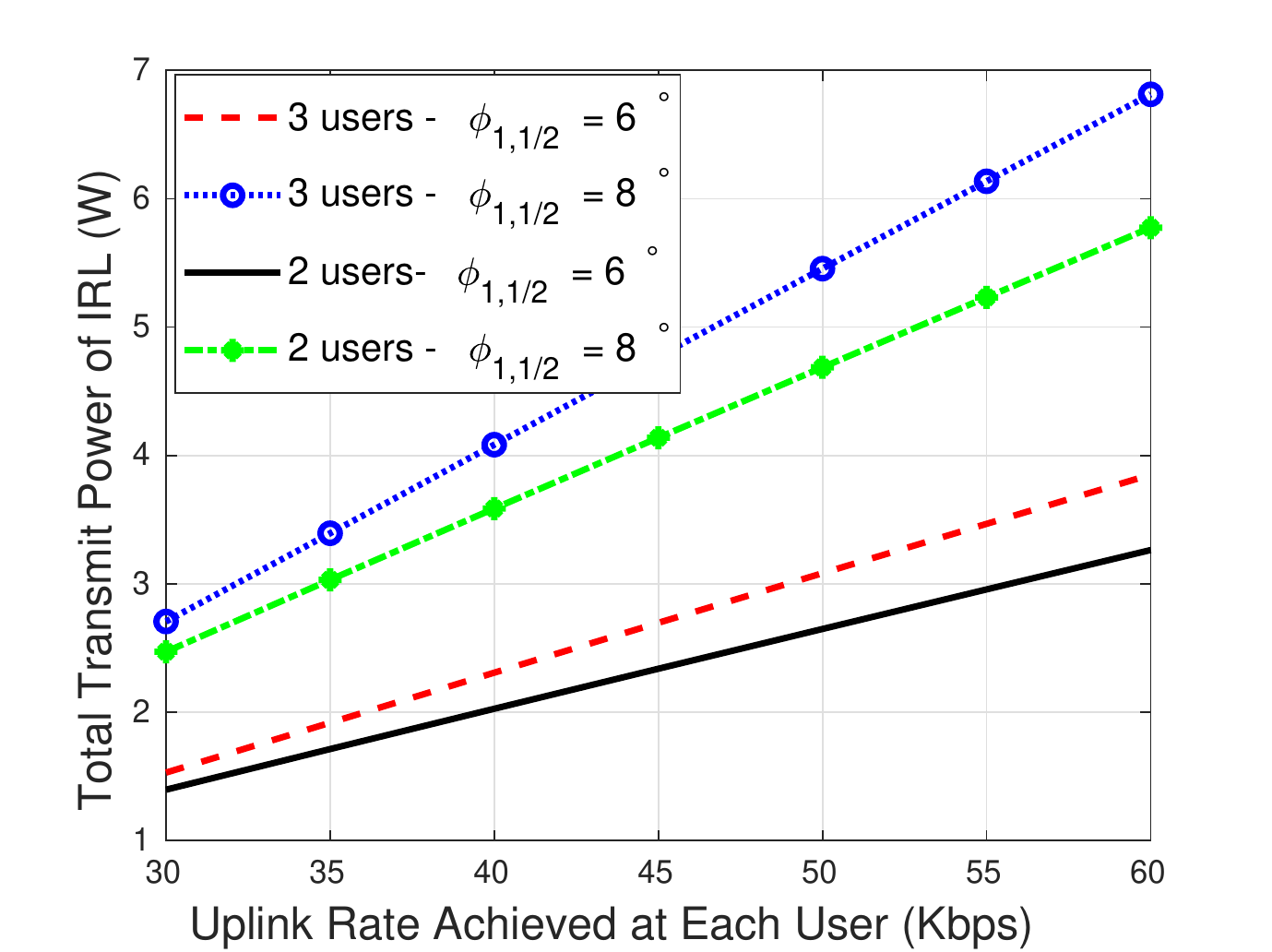}}
   \caption{
       Total transmit IRL power versus the uplink rate.
    }
    \label{fig:sim1}
\end{figure}

\begin{table}[h]
\begin{centering}
\caption{Important parameters}
\par\end{centering}
\centering{}%
\begin{tabular}{|l|c|}
\hline
\textbf{Parameters} &
\textbf{System values}
\tabularnewline
\hline
The optical band-pass filter gain, $T_s(\psi_{i,j} ) $ & 1
\tabularnewline
\hline
The field of view, $\psi_{i,j,c}$& $70^\circ$ \tabularnewline
\hline
The LED semiangle at half-power, $\phi_{0,1/2} $ & $60^\circ$ \tabularnewline
\hline
The photodetector's active area, $A_j$ & $85$ cm$^2$ (phone screens) \tabularnewline
\hline
VL transmit power, $P_{0,j}$ & $28$ W \cite{Brien} 
\tabularnewline
\hline
\end{tabular}
\end{table}

Fig. \ref{fig:sim1} presents the variation of the total transmit IRL power with respect to the uplink rate threshold $\{ \theta_j \}$ for different values of $\phi_{1,1/2}$. It can be observed that a higher uplink requires a higher IRL power that needs to be transferred by the optical transmitter. Furthermore, using a lower $\phi_{1,1/2}$ (which implies a narrower IRL beam) reduces the required transmit power while maintaining the same uplink rate. With the used system settings, the uplink rate needed for the FL updates is 36 Kbps \cite{nguyentran2019}. Therefore, the proposed approach can support the mobile devices in handling the FL tasks without expending any power from their batteries.

\begin{figure}[t]
\centering
{\includegraphics[width=0.32\textwidth]{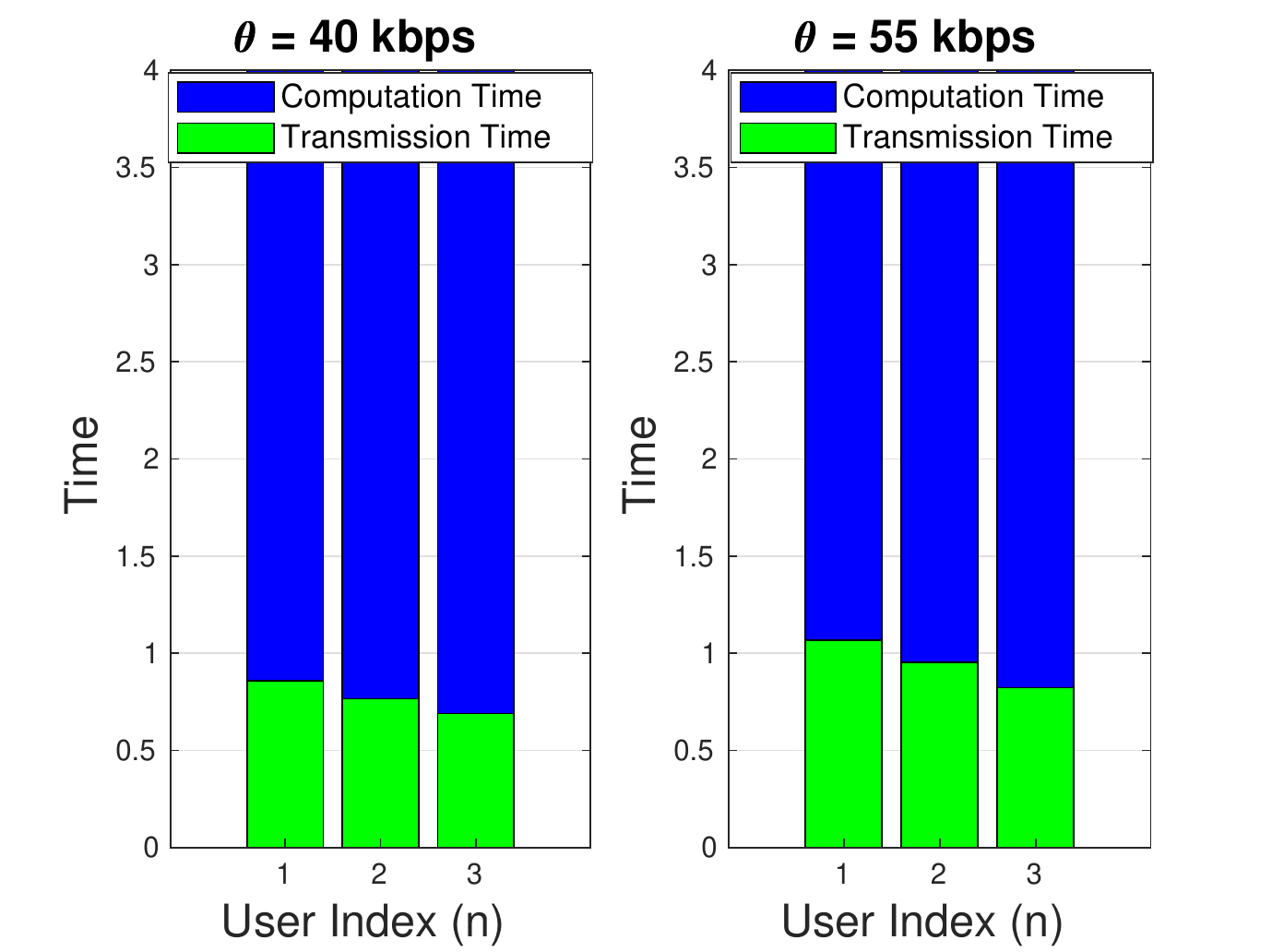}}
   \caption{
       Transmission time vs. computation time ($\{\theta_j\}=\theta$).
    }
    \label{fig:sim2}
\end{figure}

In Fig. \ref{fig:sim2}, the optimal ratios of $\{T_j^\mathrm{trans}\}$ to $\{T_j^\mathrm{comp}\}$ at the devices are shown for different values of the uplink rate. It is observed that device $3$, the nearest device to the RF AP, has the shortest transmission time. 
This can be explained the fact that a longer distance requires a higher $E_j^\mathrm{trans}$ and hence a higher $T_j^\mathrm{trans}$ following from \eqref{eq:Etrans}. Further, setting a higher $\theta$ requires, not only a higher transmit power (as in Fig. \ref{fig:sim1}), but also a longer transmission time.
In these cases, since the computation tasks are the same for all the devices, the optimal management implies increasing $\{T_j^\mathrm{trans}\}$ rather than $\{T_j^\mathrm{comp}\}$.

\begin{figure}[t]
\centering
{\includegraphics[width=0.32\textwidth]{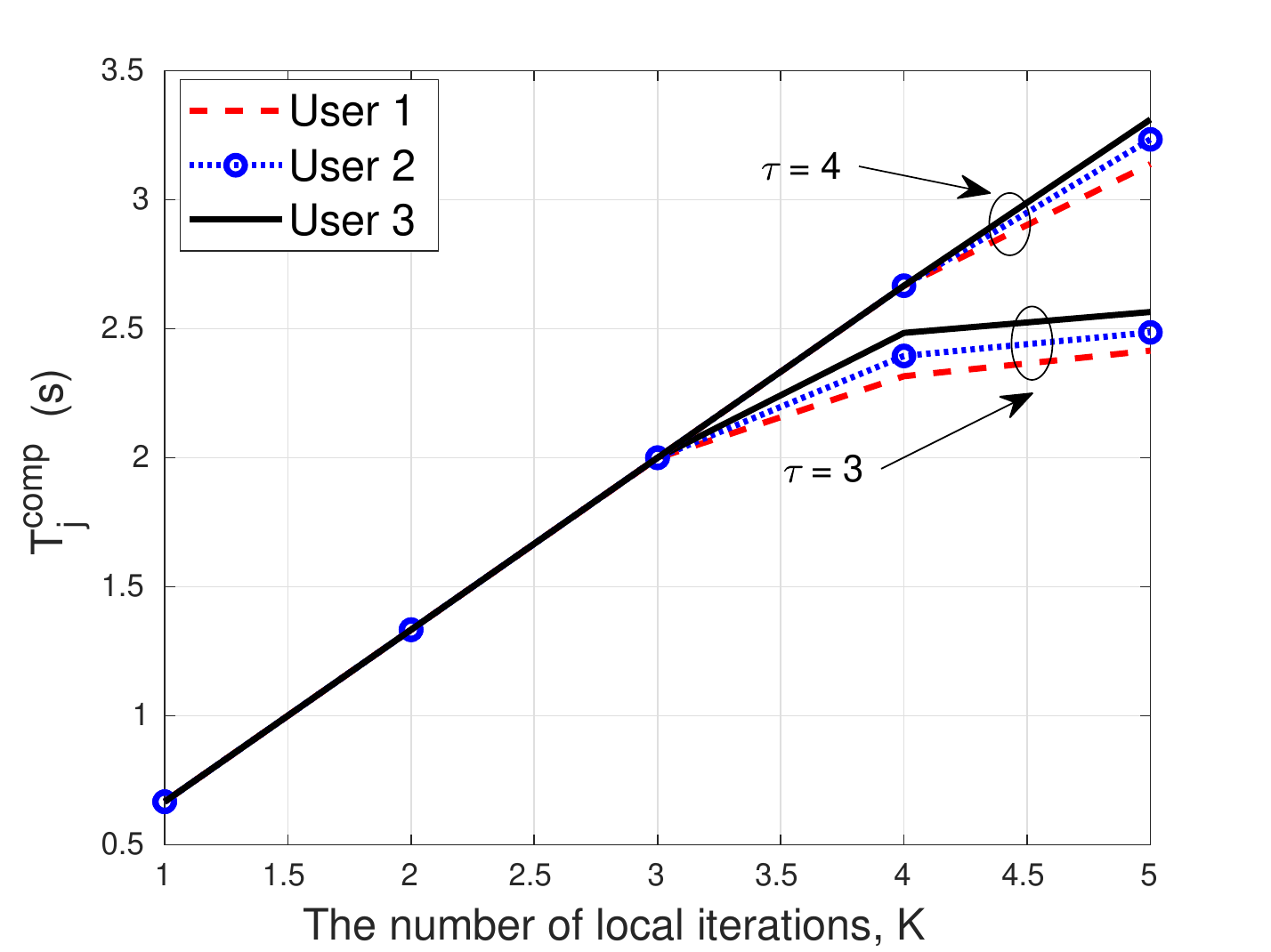}}
   \caption{
       The computation time vs. the number of local iterations.
    }
    \label{fig:sim3}
\end{figure}

\begin{figure}[t]
\centering
{\includegraphics[width=0.32\textwidth]{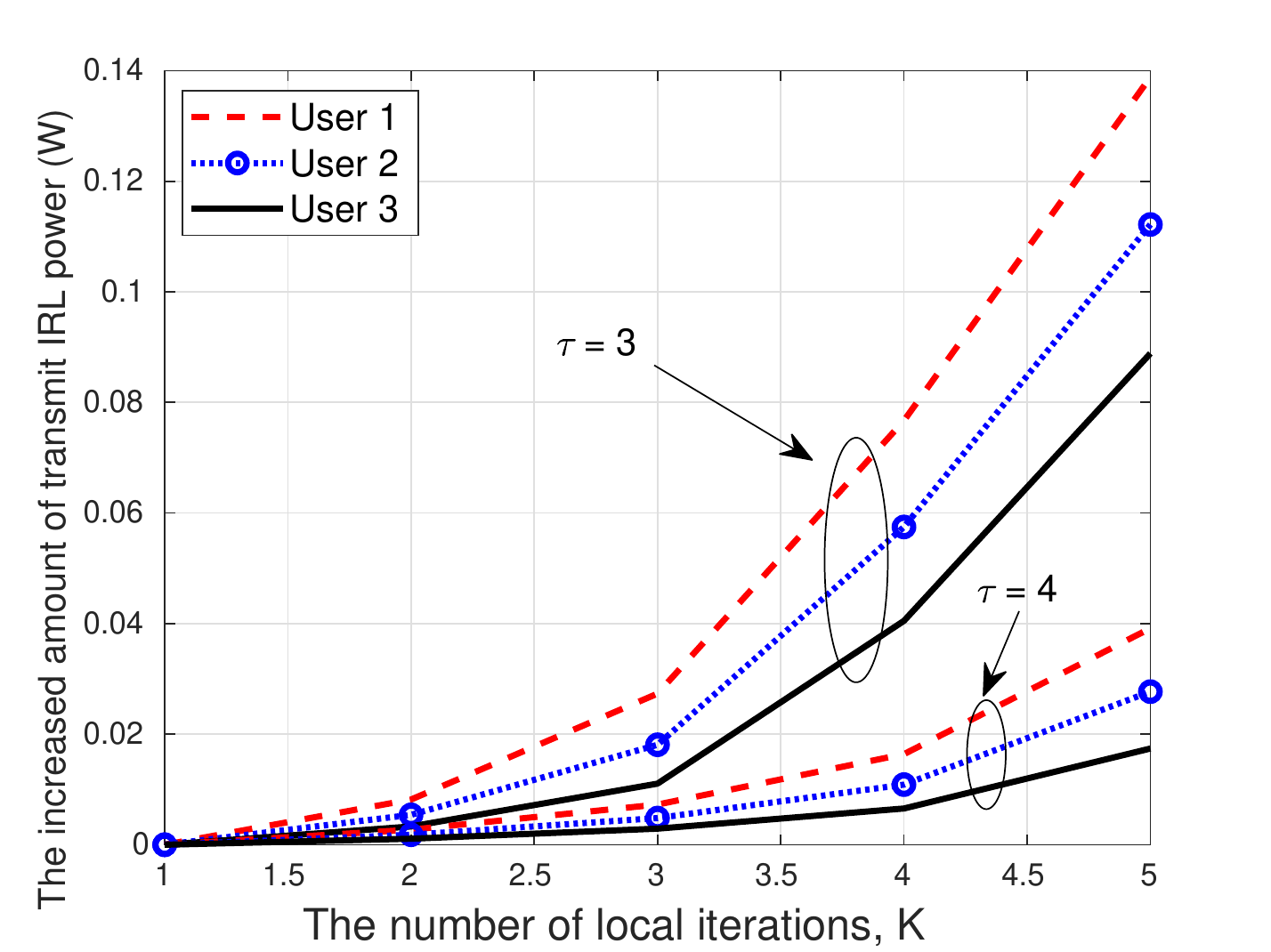}}
   \caption{
       The additional transmit power vs. the number of local iterations.
    }
    \label{fig:sim4}
\end{figure}

In Figs. \ref{fig:sim3} and \ref{fig:sim4}, the impacts of $\{ K_j\}$ on $\{T_j^\mathrm{comp}\}$ and the transmit IRL power are presented. Herein, we set $\{ K_j\} = K$, and $\theta = 40$ kbs. 
Interestingly, in Fig. \ref{fig:sim3}, the gap of $\{T_j^\mathrm{comp}\}$ between the two cases of $\tau$ increases as $K$ increases. It is due to the computation consumes more energy than the transmission. Hence, $\{T_j^\mathrm{comp}\}$ is given as much as possible to minimize the total energy consumption. Furthermore, Fig. \ref{fig:sim4} shows that increasing $\tau$ yields a lower level of additional transmit IRL power. 
This can be explained by that the CPU energy consumption refers to the energy consumed to process a certain amount of data during a time interval. If the amount of data is unchanged; a longer time interval results in a lower energy consumed for processing and a lower additional transmit IRL power.

\section{Conclusion}
In this work, we proposed for the first time the application of the lightwave power transfer to the FL-based wireless networks.
On this basis, we devised a strategy to manage the power efficiency of the network and formulated the corresponding optimization problem.
Moreover, we provided the algorithms to tackle the problem optimally.
The numerical results indicate that the proposed scenario can sufficiently replenish energy for the terminal devices to open up new opportunities for sustainable FL-based wireless networks.

\bibliographystyle{IEEEtran}
\bibliography{IEEEabrv,REF}
\end{document}